\documentclass[11pt]{article}

\usepackage[utf8]{inputenc}
\usepackage{amssymb, amsfonts, graphicx, color, subfig, ae, aecompl}
\usepackage[english]{babel}
\usepackage[intlimits]{amsmath}
\usepackage[dvipdfm]{hyperref}
\usepackage[heightrounded,a4paper]{geometry}

\newcommand{\mfb}{\mathfrak{b}}
\newcommand{\mfB}{\mathfrak{B}}
\newcommand{\mcH}{\mathcal{H}}
\newcommand{\ol}{\overline}
\newcommand{\tr}{\mathrm{tr}}

\newcommand{\mcN}{\mathcal{N}}
\newcommand{\ve}{\varepsilon}
\newcommand{\NN}{N\!N}
\newcommand{\NNN}{N\!N\!N}

\title{Quantum phase transitions and thermodynamics of quantum antiferromagnets with competing interactions}

\author{Christian Trippe\thanks{e-mail:   \href{mailto:trippe@physik.uni-wuppertal.de}{\protect\nolinkurl{trippe@physik.uni-wuppertal.de}}} \  and Andreas Kl\"umper\thanks{e-mail:   \href{mailto:kluemper@physik.uni-wuppertal.de}{\protect\nolinkurl{kluemper@physik.uni-wuppertal.de}}}\\
\parbox{0.9 \linewidth}{\vspace{0.4 \baselineskip}\centering
    Fachbereich C -- Physik, Bergische Universit\"at Wuppertal,\\
    42097 Wuppertal, Germany}}

\date{\today}

\begin{document}

\maketitle

\begin{abstract}
We study the isotropic Heisenberg chain with nearest and next-nearest neighbour interactions. The ground state phase diagram is constructed in dependence on the additonal interactions and an external magnetic field. The thermodynamics is studied by use of finite sets of non-linear integral equations resulting from integrabiliy. The equations are solved numerically and analytically in suitable limiting cases. We find second and first order transition lines. The exponents of the low temperature asymptotics at the phase transitions are determined.
\end{abstract}

\section{Introduction}

Low dimensional quantum systems are of considerable current interest. On one hand they can be studied in experiments, where they are realised as quasi 1D or 2D subsystems. On the other hand, some of the 1D systems, like the spin-$\frac{1}{2}$ Heisenberg chain, can be solved exactly or otherwise non-perturbativly. The method of exactly solving quantum spin systems via Bethe ansatz is essentially restricted to 1D models, but allows for solving models of coupled chains, see \cite{zvyagin:BA-multichain} and references therein.
Depending on the topology, coupled chains (or spin ladders) may be considered as interpolations between 1D and 2D systems, or in the case of spin ladders with zigzag interactions the system may be viewed as a single chain with longer range interactions.

In many cases these models are only studied in the ground state and without an external magnetic field, see e.g.~\cite{MuellerVekuaMikeska2002}. Considering nonzero temperature and an external magnetic field is of interest for two reasons. First, the magnetic field can lead to (several) quantum phase transitions in the ground state, see e.g.~\cite{FrahmRoed:spinladder99}. Second, nonzero temperature and magnetic field are required for comparison with experimental work.

The main goal of this paper is to study the thermodynamical properties of two quantum spin chains with competing interactions in an external magnetic field. Both are generalisations of the standard spin-$\frac{1}{2}$ Heisenberg chain and can be solved exactly via Bethe ansatz \cite{ZvyKl:qpt}. Here we investigate in more detail the thermodynamics which also leads to new information on the ground state.

The paper is organized in the following way. First we introduce in sect.~\ref{sec:hamiltonians} the Hamiltonians of the two models investigated in this paper. Then in sect.~\ref{sec:transfermatrices} we show the relation of these Hamiltonians to a row-to-row transfer matrix and present the definition of a quantum transfer matrix which allow for exactly solving the models via Bethe ansatz. In sect.~\ref{sec:nlie} we derive non-linear integral equations determining the thermodynamical properties of the models and consider their zero temperature limit. Using these equations we discuss the ground state phase diagrams in sect.~\ref{sec:phase-diagram} and present several results for the magnetic susceptibility, magnetisation and specific heat in sect.~\ref{sec:thermodynamics}. Finally, we summarise our results.

\section{Hamiltonians}
\label{sec:hamiltonians}

In this paper we investigate the properties of two systems. First, the Hamiltonian of the Bethe ansatz solvable isotropic spin-$\frac{1}{2}$ chain with nearest neighbour interactions and three spin interactions between successive spins, referred to as next-nearest neighbour model or ``model with NN interaction'' \cite{ZvyKl:qpt,Frahm:xxz,Tsvelik:mxxz}, can be written in the form \mbox{$\mcH_{\NN}=J\mcH_1+\alpha_2J^2\mcH_2+\mcH_h$}, where
\begin{equation}
\mcH_1=\sum_{i=1}^L \left( \vec{S}_i \vec{S}_{i+1} - \frac{1}{4} \right)
\end{equation}
is the Hamiltonian of the standard spin-$\frac{1}{2}$ Heisenberg chain, and
\begin{equation}
\mcH_2=\sum_{i=1}^L \vec{S}_i \cdot \left( \vec{S}_{i-1} \times \vec{S}_{i+1} \right )
\end{equation}
contains the three spin interactions. The Zeeman term
\begin{equation}
\label{eq:mcH_h}
\mcH_h=-h\sum_{i=1}^LS_i^z
\end{equation}
takes account of an homogeneous external magnetic field $h$. The coupling $\alpha_2$ determines the relative strength of the three-spin interactions.

The second system considered in this paper contains nearest and next-nearest neighbour interactions as well as four-spin interactions (in the following referred to as ``model with NNN interactions" \cite{ZvyKl:qpt,MuramotoTakahashi:four-body}) which has the form $\mcH_{\NNN}=J\mcH_1+\alpha_3J^3\mcH_3+\mcH_h$ where
\begin{equation}
\label{eq:mcH_3}
\mcH_3=\sum_{i=1}^L \left [ -\vec{S}_i\vec{S}_{i+1} + \frac{1}{2}\vec{S}_i\vec{S}_{i+2} + 2\left (\vec{S}_{i-1}\vec{S}_{i+1}\right )\left (\vec{S}_i\vec{S}_{i+2}\right ) -2 \left (\vec{S}_{i-1}\vec{S}_{i+2}\right )\left (\vec{S}_i\vec{S}_{i+1}\right ) +\frac{1}{8} \right ]\;.
\end{equation}
The operators $\mcH_i$, ($i=1,2,3$) and $\mcH_h$ commute mutually as well as with a transfer matrix $t(\lambda)$ constructed in the next section. These properties allow for exactly solving the models via Bethe ansatz.

\section{Transfer matrices}
\label{sec:transfermatrices}

The $R$-matrix belonging to the Heisenberg chain is given by
\begin{equation}
\label{eq:r-matrix}
R^{\alpha\gamma}_{\beta\delta}(\lambda,\mu)=\delta^\alpha_\delta \delta^\gamma_\beta+(\lambda-\mu)\delta^\alpha_\beta \delta^\gamma_\delta
\end{equation}
which is a solution of the Yang-Baxter equation. Here the indices in the first column denote states in the auxiliary space and the indices in the second column denote states in the quantum space. For a more detailed description of the notation see \cite{GK:hubbard-model}. With $(L_j)^\alpha_\beta(\lambda,\mu)=R^{\alpha\gamma}_{\beta\delta}(\lambda,\mu){e_j}_{\gamma}^{\delta}$ the operator defined on a chain of $L$ sites
\begin{equation}
t(\lambda)=\tr\left[L_L(\lambda,0)\dotsm L_1(\lambda,0)\right]
\end{equation}
yields a family of commuting transfer matrices $[t(\lambda),t(\mu)]\!=\!0$ for arbitrary $\lambda, \mu \in \mathbb{C}$. The operators ($\mcH_i$, $i=1,2,3$) are given as logarithmic derivatives of this transfer matrix at the shift point $\lambda=0$
\begin{eqnarray}
\label{eq:mch-tau-1}
\mcH_1&=&\frac{1}{2}\tau^{(1)}(0)-\frac{L}{2} \label{eq:mcH_1-tau}\\
\label{eq:mch-tau-2}
\mcH_2&=&\frac{\imath}{4}\tau^{(2)}(0)+\frac{\imath L}{4} \label{eq:mcH_2-tau}\\
\label{eq:mch-tau-3}
\mcH_3&=& -\frac{1}{8}\tau^{(3)}(0)+\frac{L}{4} \label{eq:mcH_3-tau}
\end{eqnarray}
where $\tau (\lambda)=\ln t(\lambda)$.

Next, we introduce $R$ matrices $\ol{R}^{\alpha\gamma}_{\beta\delta}(\lambda,\mu)=R^{\gamma\beta}_{\delta\alpha}(\mu,\lambda)$ by a clockwise and $\ol{R}^{\alpha\gamma}_{\beta\delta}(\lambda,\mu)=R^{\gamma\beta}_{\delta\alpha}(\mu,\lambda)$ by an anticlockwise rotation and define a transfer matrix $\ol{t}(\lambda)$ in the same way as $t$ above. The partion function of the models without magnetic field can be expressed as
\begin{equation}
\label{eq:zustandssumme}
Z=\lim_{N\rightarrow\infty} \tr \left[\prod_{i=1}^{N/2}t(u_i)\ol{t}(0)\right]
\end{equation}
with appropriate spectral parameters $u_i$ \cite{KlSa:thermal}, depending on the model.\footnote{Note that we ignored in \eqref{eq:zustandssumme} the additive constants of equations \eqref{eq:mch-tau-1}-\eqref{eq:mch-tau-3}.} The column-to-column transfer matrix of the corresponding two dimensional $L\times N$ lattice is called quantum transfer matrix (QTM). It is defined by
\begin{equation}
t^{QTM}(\lambda)=\tr \left(DL_N^{QTM}(\lambda,0)L_{N-1}^{QTM}(\lambda,u_{N/2})\dotsm L_2^{QTM}(\lambda,0) L_1^{QTM}(\lambda,u_1)\right)
\end{equation}
where the magnetic field $h$ is included by means of twisted boundary conditions via the diagonal matrix $D=\mathrm{diag}\left(\exp(\beta h/2),\exp(-\beta h/2)\right)$ and 
\begin{equation}
{L_j^{QTM}}^\alpha_\beta(\lambda,\mu)=\begin{cases}
R^{\alpha\gamma}_{\beta\delta}(\lambda,\mu){e_j}^\delta_\gamma\,, & j \text{ even}\\
\tilde{R}^{\alpha\gamma}_{\beta\delta}(\lambda,\mu){e_j}^\delta_\gamma=R^{\delta\alpha}_{\gamma\beta}(\mu,\lambda){e_j}^\delta_\gamma\,, & j \text{ odd}\,.
\end{cases}
\end{equation}
The monodromy matrix corresponding to the QTM is a representation of the Yang-Baxter algebra with intertwiner $R$. The partion function is given by
\begin{equation}
Z_N=\tr\left(t^{QTM}(0)\right)^L\;.
\end{equation}
Hence the free energy in the thermodynamic limit is determined by
\begin{align}
\label{eq:freie-energie-eigenwert}
f=-T\lim_{N\rightarrow\infty}\ln \Lambda^{QTM}(0)
\end{align}
where $\Lambda^{QTM}$ is the largest eigenvalue of the QTM.

\section{Non-linear integral equations}
\label{sec:nlie}

For the standard spin-$\frac{1}{2}$ Heisenberg chain one can derive different sets of non-linear integral equations (NLIE) determining the thermodynamical properties.
Historically first, an infinite set of equations via TBA \cite{Takahashi:TBA} was obtained. Then, as a second possibility, a set of only two equations \cite{Kluemper92,Kluemper93} was derived. In fact, it is also possible to find an arbitrary number of equations interpolating between these extreme schemes \cite{Su:fh}.

For the models investigated in this paper the set of two coupled non-linear integral equations was derived in \cite{ZvyKl:qpt} where also certain parameter ranges were treated numerically. Here, we performed further numerical studies of these equations and found that they are not valid for low temperatures in the vicinity of the phase coexistence if straight integration contours are used (see sect.~\ref{sec:phase-diagram}). This is due to the fact that the imaginary parts of some Bethe ansatz numbers grow strongly, leading to a crossing of the integration contours by singularities of the integrands.

To determine the free energy also for the cases where the two NLIE with standard contours are not valid, it is useful to utilise the fusion hierarchy of this model.
One obtains in the usual way an infinite set of NLIE.
The first equation is
\begin{equation}
\label{eq:y1-NN}
\ln y_1(x)=-\frac{v\beta}{\cosh (\pi x)}-Z(x)+\left(s\ast \ln Y_2\right)(x)
\end{equation}
with $\ast$ denoting convolution\footnote{$\left(g \ast h \right ) (t) =  \int_{-\infty}^\infty g (t- \tau)\,h(\tau)\,d\tau$} and $s(x)=\dfrac{1}{2\cosh(\pi x)}$ being the integration kernel. $Z(x)$ depends on the model and is given by
\begin{equation}
Z(x)=\begin{cases}
      \alpha_2v^2\beta\, \dfrac{\sinh (\pi x)}{\cosh^2 (\pi x)}\,, & \text{for the model with NN interactions}\,,\\
	\alpha_3v^3\beta\, \dfrac{2\tanh^2 (\pi x) -1}{\cosh (\pi x)}\,, & \text{for the model with NNN interactions}\,.
     \end{cases}
\end{equation}
The other equations are independent of the model and read
\begin{equation}
\ln y_j(x)=\left(s\ast \ln\left(Y_{j-1}Y_{j+1}\right)\right)(x)\;.
\end{equation}
The magnetic field does not enter explicitly in these equations, it only fixes the asymptotic behaviour of the $y$-functions. For zero magnetic field it reads $\lim_{|x|\rightarrow\infty}y_j(x)=j(j+2)$ and for $h\neq0$
\begin{equation}
\label{eq:y-asymptotic}
\lim_{|x|\rightarrow\infty}y_j(x)=\left(\frac{z^{j+1}-z^{-(j+1)}}{z-z^{-1}}\right)^2-1
\end{equation}
with $z=e^{\beta h/2}$. Note that the asymptotic behaviour and therefore the system of equations is invariant under a change of sign of the magnetic field.
The free energy is given by
 \begin{equation}
\label{eq:freie-energie-fusion}
f=e_0-T\int_{-\infty}^\infty dx \, \frac{\ln Y_1(x)}{2\cosh (\pi x)}
\end{equation}
with $e_0=-J\ln2$ for the model with NN interactions and $e_0=-J\ln2+\alpha_3 \frac{3J^3}{8}\zeta (3)$ for the model with NNN interactions, $\zeta$ denoting the Riemann $\zeta$-function.

It is possible to close the set of infinitely many integral equations after the $(k-1)$th equation. The minimal paramter $k=1$ gives the set of two NLIE presented in \cite{ZvyKl:qpt}. Next we show the results for $k \geq2$ following \cite{Su:fh}.

One can find suitable functions $\mfb$, $\ol{\mfb}$, $\mfB(x):=1+\mfb(x)$ and $\ol{\mfB}(x):=1+\ol{\mfb}(x)$ satisfying $\mfB(x)\ol{\mfB}(x)=Y_k(x)$ and therefore yielding the functional relation
\begin{equation}
\label{eq:functionalbeziehung_y}
y_{k-1}(x+\frac{\imath}{2})y_{k-1}(x-\frac{\imath}{2})=Y_{k-2}(x)\mfB(x)\ol{\mfB}(x)\;.
\end{equation}
For \eqref{eq:functionalbeziehung_y} the cases $k\!=\!1$ and $k\!=\!2$ are exceptional. For $k\!=\!1$ the equation is not used, for $k\!=\!2$ we use $Y_0\equiv1$. This leads to the following NLIE
\begin{equation}
\label{eq:lny_k}
\ln y_{k-1}(x)=s\ast \ln\left(Y_{k-2}\right)(x)+s\ast \ln\left(\mfB\ol{\mfB}\right)(x)\;.
\end{equation}
Finally, two NLIE equations for $\mfb$, $\ol{\mfb}$ close the equation system exactly
\begin{align}
\label{eq:lnb}
\ln \mfb(x)&=\frac{\beta h}{2} + \left(s\ast \ln Y_{k-1}\right)(x)+\left(\kappa \ast \ln \mfB\right)(x)-\left(\kappa \ast \ln \ol{\mfB}\right)(x+\imath)\\
\label{eq:lnb_quer}
\ln \ol{\mfb}(x)&=-\frac{\beta h}{2} + \left(s\ast \ln Y_{k-1}\right)(x)+\left(\kappa \ast \ln \ol{\mfB}\right)(x)-\left(\kappa \ast \ln \mfB\right)(x-\imath)\;.
\end{align}
In \eqref{eq:lny_k} for $k\!=\!2$ and in \eqref{eq:lnb}, \eqref{eq:lnb_quer} for $k\!=\!1$, instead of $s\ast \ln Y_0$ the inhomogeneity of \eqref{eq:y1-NN} has to be used. The kernel function $\kappa$ is given by
\begin{equation}
\label{eq:kern-kappa}
\kappa(x)=\frac{1}{2\pi}\int_{-\infty}^\infty dk\, \frac{e^{-|k|/2}}{2\cosh (k/2)}\,e^{\imath kx}\;.
\end{equation}
Here, the magnetic field enters the equations explicitly.
The different sets of equations for different $k$ are all equivalent. The equations for larger $k$ allow for the investigation of lower temperatures even with straight contours.

Calculating the zero temperature limit of the fusion hierarchy with $\ve_i=T\ln y_i$ in the usual way \cite{Takahashi:TBA} one obtains only one dressed energy $\ve:=\ve_1$ which can have negative values, all other dressed energies are strictly positive and hence do not contribute. The dressed energy $\ve$ is determined by
\begin{equation}
\label{eq:dressedenergy2}
\ve(\lambda)+\frac{1}{\pi}\int_\mcN d\mu\, \frac{1}{1+(\lambda-\mu)^2}\,\ve(\mu)=\ve_0(\lambda)
\end{equation}
with
\begin{equation}
\label{eq:bareenergy2}
\ve_0(\lambda)=-\frac{2J}{1+4\lambda^2}+
\begin{cases}
-\frac{8\alpha J}{\pi} \frac{2\lambda}{(1+4\lambda^2)^2}+|h|\,, & \text{ for } \mcH_{NN} \\
-\frac{16\alpha J}{\pi^2} \frac{12\lambda^2-1}{(1+4\lambda^2)^3}+|h|\,, & \text{ for } \mcH_{NNN}
\end{cases}
\end{equation}
and $\mcN=\{ \mu\in\mathbb{R}|\ve(\mu)<0\}$. Note that only the absolute value of the magnetic field enters the equation \eqref{eq:dressedenergy2} via the bare energy \eqref{eq:bareenergy2}. This is due to the fact that only the absolute value of $h$ enters the fusion hierarchy via the asymptotic behaviour \eqref{eq:y-asymptotic}. This statement becomes even more obvious if one derives \eqref{eq:dressedenergy2} from the system of only two NLIE. Here one of the steps of the derivation uses the observation that the function $\ol{\mfb}$ drops out for $h>0$ and $\mfb$ drops out for $h<0$ and in both cases \eqref{eq:dressedenergy2} is obtained. Furthermore a rescaling of the couplings $\alpha_2$, $\alpha_3$ was applied by $\alpha=\alpha_2 v$ for the model with NN interactions and $\alpha=\alpha_3 v^2$ for the model with NNN interactions. This will be useful in the discussion of the ground state phase diagram in the next section as in the new variable certain critical points occur at $\alpha=\pm 1$.

The dressed energy can be solved in two limiting cases. The first one is at $h\!=\!0$ and $|\alpha| \leq 1$ by Fourier transform
\begin{equation}
\label{eq:dressedenergy-h=0}
\ve(\lambda)=
\begin{cases}
-\frac{v}{\cosh (\pi \lambda)}\left[1+\alpha\tanh (\pi \lambda) \right] & \text{ for } \mcH_{\NN}\,, \\
-\frac{v}{\cosh (\pi \lambda)}\left[1-\alpha+2\alpha \tanh^2 (\pi \lambda) \right] & \text{ for } \mcH_{\NNN}\,.
\end{cases}
\end{equation}
The second case is for $h \geq h_f$ where the integral vanishes and $\ve(\lambda)=\ve_0(\lambda)$. $h_f$ is the saturation field, i.e.~the value of the magnetic field corresponding to the phase transition into the ferromagnetic phase.

\section{Phase diagrams of the ground state}
\label{sec:phase-diagram}

\begin{figure}[h!t]
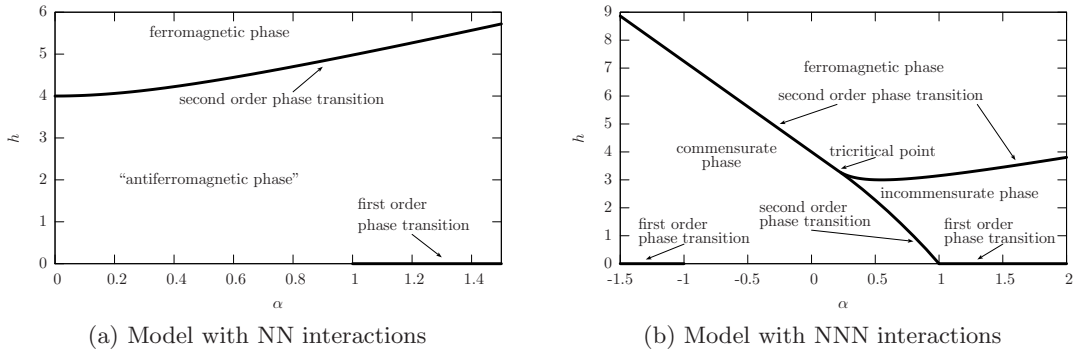

\centering
\subfloat[][Model with NN interactions]{\label{fig:NN-phasendiagramm-ende} \resizebox{7cm}{!}{\input phasendiagramm_NN_ende_english.tex}}\quad
\subfloat[][Model with NNN interactions]{\label{fig:NNN-phasendiagramm-ende} \resizebox{7cm}{!}{\input phasendiagramm_NNN_ende_english.tex}}
\caption{Phase diagrams of the model with NN and NNN interactions. Both phase diagrams are symmetric with respect to a change of sign of the magnetic field $h$, and for the model with NN interactions \protect\subref{fig:NN-phasendiagramm-ende} also with respect to a change of sign of the coupling $\alpha$. In these pictures the normalization $J\!=\!2$ in $\mcH_{N\!N}$ and $\mcH_{N\!N\!N}$ is used. The first order phase transitions are exactly at $h\!=\!0$.}
\label{fig:phasendiagramm-ende}
\end{figure}

In this section we first give our results on the ground state phase diagram for both models (Fig.~\ref{fig:phasendiagramm-ende}). Our results differ a little from those of \cite{ZvyKl:qpt}. For all couplings $\alpha$ there is a phase transition into a ferromagnetic phase and phase coexistence for the lines $|\alpha|\geq 1$, $h=0$. But only in the model with NNN interactions and positive coupling $\alpha$ we find a phase transition between a commensurate and an incommensurate phase.

\begin{figure}[t]
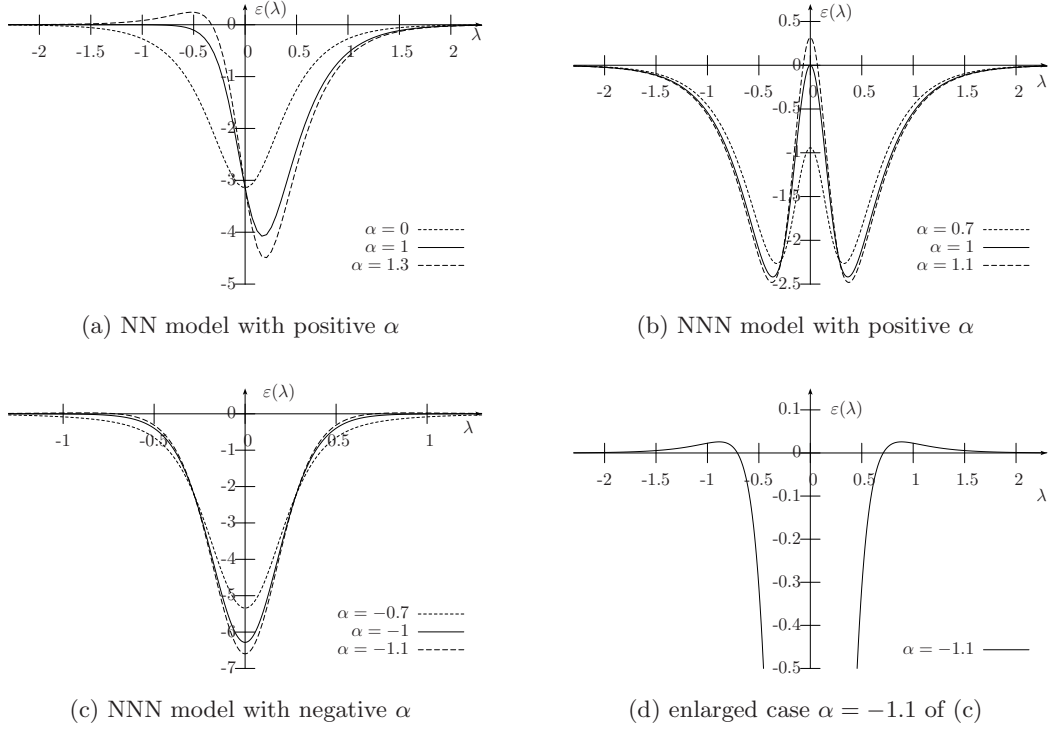

\centering
\subfloat[][NN model with positive $\alpha$]{\label{fig:NN-dressedenergy} \resizebox{7cm}{!}{\input dressed-energy-NN-diverse.tex}}\quad
\subfloat[][NNN model with positive $\alpha$]{\label{fig:NNN-dressedenergy} \resizebox{7cm}{!}{\input dressed-energy-NNN-diverse.tex}}\\
\subfloat[][NNN model with negative $\alpha$]{\label{fig:NNN-dressedenergy2} \resizebox{7cm}{!}{\input dressed-energy-NNN-diverse2.tex}}\quad
\subfloat[][enlarged case $\alpha=-1.1$ of \protect\subref{fig:NNN-dressedenergy2}]{\label{fig:NNN-dressedenergy-klein} \resizebox{7cm}{!}{\input dressed-energy-NNN-a=-1,1-fein-label.tex}}\\
\caption{The dressed energies according to equation \eqref{eq:dressedenergy-h=0} for $h=0$ and normalisation $J\!=\!2$. The data are approximate for $|\alpha|>1$.}
\label{fig:dressedenergy}
\end{figure}

Looking at the dressed energy for the models with zero magnetic field one can understand the phase diagrams. The phase transitions in dependence on the magnetic field $h$ correspond to the opening and closing  of Fermi seas, i.e.~appearance or disappearance of intervals of negative energy modes of the dressed energy. Qualitatively, these transitions occur at magnetic fields $h$ coinciding with extremal values of the dressed energy at zero field as plotted in Fig.~\ref{fig:dressedenergy}. This is very much like the discussion of van-Hove singularities of free particle systems. However, here we deal with an interacting system for which the chemical potential is not identical to $h$ but equal to $|h|$. So only the negative extremal values of the dressed energy are relevant.
Although the analytic solution \eqref{eq:dressedenergy-h=0} is strictly valid only for $|\alpha|\leq1$ we use these formulas for slightly larger values of $|\alpha|$ where they should be good approximations to the true solution.

Depending on the longer range coupling $\alpha$, the model with NNN interactions has one or two Dirac seas. For $\alpha \lesssim 0.2$ only a second order phase transition into the ferromagnetic phase and a first order transition line with non vanishing spontaneous magnetisation for $\alpha \leq -1$, $h=0$ exist (Fig.~\ref{fig:NNN-dressedenergy2} and \ref{fig:NNN-dressedenergy-klein}). For $\alpha \gtrsim 0.2$  (Fig.~\ref{fig:NNN-dressedenergy}) the dressed energy possesses two local minima and one local maximum. The two local minima have identical value as $\ve(\lambda)$ is an even function. So here a phase transition between two phases exists (denoted ``commensurate'' and ``incommensurate''). The tricitical point can be determined to $\alpha=\frac{\pi^2}{48}\simeq 0.2056$ and $h_f=\frac{5J}{3}$ by analysing the solution of the dressed energy in the ferromagnetic phase.

The phase diagram of the model with NN interactions is symmetric under a sign change of the coupling $\alpha$, as in the NLIE this can be compensated for by a sign change of the spectral parameter $\lambda$. Hence for the model with NN interactions it is sufficient to look at couplings $\alpha \geq 0$. This model has always only one Dirac sea (Fig.~\ref{fig:NN-dressedenergy}), hence there is only a second order phase transition between the antiferromagnetic and the ferromagnetic phase and a  first order transition line with non vanishing spontaneous magnetisation for $|\alpha| > 1$, $h=0$.

The critical field corresponding to the phase transition from the antiferromagnetic into the ferromagnetic phase can be determined as usual \cite{KBI:qism} from the bare energy \eqref{eq:bareenergy2}. For the model with NNN interactions one obtains
\begin{equation}
\label{eq:h_f}
h_f=
\begin{cases}
2J-\alpha \frac{16}{\pi^2}J\,, & \text{for } \alpha \le \pi^2/48\,,\\
\frac{2J}{\sqrt{24\frac{\alpha}{\pi^2}\left(24\frac{\alpha}{\pi^2}+4\right)}-24\frac{\alpha}{\pi^2}}+16J\frac{\alpha}{\pi^2}\frac{3\sqrt{24\frac{\alpha}{\pi^2}\left(24\frac{\alpha}{\pi^2}+4\right)}-4-72\frac{\alpha}{\pi^2}}{\left(\sqrt{24\frac{\alpha}{\pi^2}\left(24\frac{\alpha}{\pi^2}+4\right)}-24\frac{\alpha}{\pi^2}\right)^3}\,, & \text{for } \alpha \geq \pi^2/48\,.
\end{cases}
\end{equation}
For $\alpha \le \pi^2/48$ this is a linear relation between the critical field $h_f$ and the coupling $\alpha$. For $\alpha \geq \pi^2/48 $ the relation is non-linear, however with linear asymptotic behaviour for large values of $\alpha$
\begin{equation}
h_f=J\left(1+4\frac{\alpha}{\pi^2}\right)+O\left(\frac{1}{\alpha}\right)\;.
\end{equation}
The applicability of \eqref{eq:h_f} is restricted to really large values of $\alpha$, e.g.~the error gets smaller than 1\% for $\alpha \simeq 5.8$. For the model with NN interactions the equation determining the minimum of the bare energy is cubic, whereas it is biquadratic for the model with NNN interactions. For this reason we want to give only numerical values for the critical field $h_f$ for the model with NN interactions. These can be taken from Fig.~\ref{fig:NN-phasendiagramm-ende}.

\section{Magnetic susceptibility, magnetisation and specific heat}
\label{sec:thermodynamics}

In this section we present the magnetic susceptibility $\chi$ for typical values of the coupling $\alpha$ in dependence on the magnetic field and temperature. We also show evidence that the phase transition at $h\!=\!0$ is of first order. Finally, we calculate the specific heat $c$ at the phase transitions. The derivatives of the free energy are obtained by differentiating \eqref{eq:freie-energie-fusion} and deriving new integral equations for the logarithmic derivatives of the auxiliary functions involving the auxiliary functions as external parameters.

As the situation in the model with NN interactions is very similar to the one for the model with NNN interactions and negative coupling $\alpha$, we will focus in the following on the model with NNN interactions and only sometimes give comments on the model with NN interactions. For the numerical calculations we always use the normalisation $J\!=\!2$.

\begin{figure}[t]
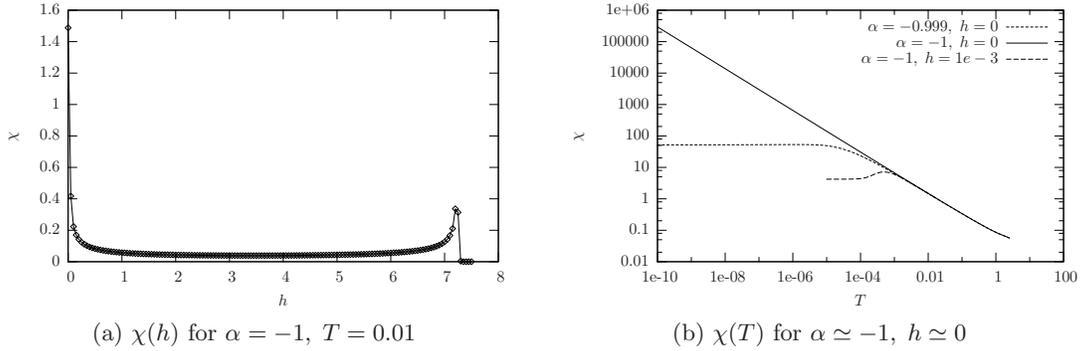

\centering
\subfloat[][$\chi(h)$ for $\alpha=-1,\;T=0.01$]{\label{fig:NNN-chi-scan-komplett-alpha-10} \resizebox{7cm}{!}{\input mag-sus_NNN_a=-1,0_komplett_neu.tex}}\quad
\subfloat[][$\chi(T)$ for $\alpha\simeq-1,\;h\simeq0$]{\label{fig:NNN-chi-scan-alpha-um-1} \resizebox{7cm}{!}{\input mag-sus_NNN_a=-1,0_drumherum.tex}}
\caption{The magnetic susceptibility for the model with NNN interactions and negative coupling $\alpha\simeq-1$.}
\label{fig:NNN-chi-h-}
\end{figure}

For the model with NNN interactions the magnetic susceptibility is shown for typical values of $\alpha$. In Fig.~\ref{fig:NNN-chi-scan-komplett-alpha-10} $\chi(h)$ is shown for $\alpha\!=\!-1$ and $T\!=\!0.01$. One sees that there is a maximum at $h\simeq7.2$ corresponding to the phase transition between the antiferromagnetic and ferromagnetic phase and another maximum at $h\!=\!0$ also corresponding to a divergence at $T\!=\!0$ (Fig.~\ref{fig:NNN-chi-scan-alpha-um-1}). This qualitative picture is also true for $\alpha<-1$.

For larger values of $\alpha$ the low field maximum exists until $\alpha\! \lesssim\! -0.2$ and $h\ll h_f$ but it does not correspond to a phase transition because the magnetic susceptibility does not diverge for $T\rightarrow 0$.

\begin{figure}[t]
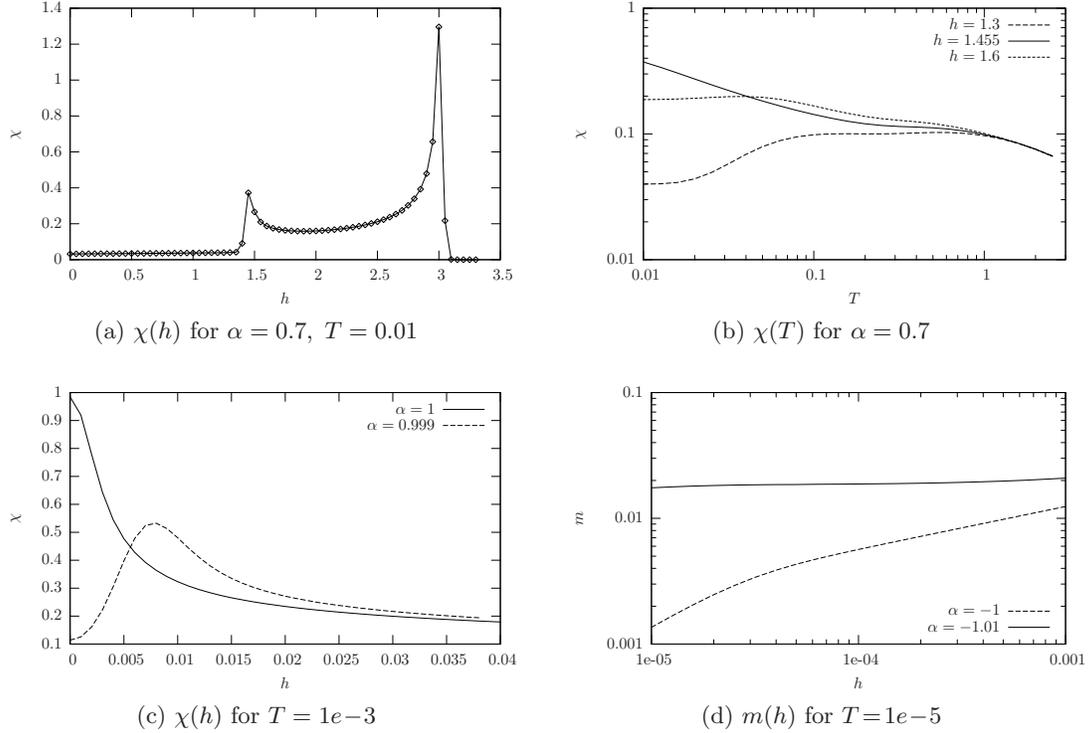

\centering
\subfloat[][$\chi(h)$ for $\alpha=0.7,\;T=0.01$]{\label{fig:NNN-chi-scan-komplett-alpha07} \resizebox{7cm}{!}{\input mag-sus_NNN_a=0,7_komplett_neu.tex}}\quad
\subfloat[][$\chi(T)$ for $\alpha=0.7$]{\label{fig:NNN-chi-alpha07-unten} \resizebox{7cm}{!}{\input mag-sus_NNN_a=0,7-unten_paper_neu.tex}}\\
\subfloat[][$\chi(h)$ for $T=1e\!-\!3$]{\label{fig:NNN-chi-scan-alpha-um1} \resizebox{7cm}{!}{\input mag-sus_NNN_um1.tex}}\quad
\subfloat[][$m(h)$ for $T\!=\!1e\!-\!5$]{\label{fig:NNN-magnetisierung-scanlog-um-1} \resizebox{7cm}{!}{\input magnetisierung_scanlog_NNN_um-1.tex}}
\caption{The magnetic susceptibility for the model with NNN interactions in figures \protect\subref{fig:NNN-chi-scan-komplett-alpha07}-\protect\subref{fig:NNN-chi-scan-alpha-um1} and the magnetisation in figure \protect\subref{fig:NNN-magnetisierung-scanlog-um-1}.
Fig.~\protect\subref{fig:NNN-chi-alpha07-unten} clearly shows that for $\alpha\!=\!0.7$ the magnetic susceptibility diverges for $h\!=\!1.455$, which is in contrast to the cases with lower and higher magnetic fields.}
\label{fig:NNN-chi-h+}
\end{figure}

For $\alpha \gtrsim 0.206$, the value corresponding to the tricritcal point, two maxima exist (Fig.~\ref{fig:NNN-chi-scan-komplett-alpha07}). Here also the maximum at lower magnetic field diverges for $T\rightarrow 0$ and so corresponds to a phase transition (Fig.~\ref{fig:NNN-chi-alpha07-unten}).
The value of the lower critical field decreases with increasing $\alpha$ and turns 0 at $\alpha\!=\!1$. This fact, known from the functional behaviour of the dressed energy is also supported by the numerical data for the magnetic susceptibility at finite temperature. In Fig.~\ref{fig:NNN-chi-scan-alpha-um1} one clearly sees that the maximum of the magnetic susceptibility occurs at a finite magnetic field for $\alpha<1$.

The value of the magnetic field corresponding to the phase transition between the commensurate and incommensurate phase could not be determined analytically. However, good numerical results are obtained by calculating the magnetic susceptibility at finite but low temperature for different values of the magnetic field and determining the local maximum. Doing this for different temperatures one can also estimate the error. The line between the commensurate and incommensurate phase in Fig.~\ref{fig:NNN-phasendiagramm-ende} is located in this way, where errorbars are within linewidth.

The magnetisation for $h\rightarrow0$ is shown for the model with NNN interactions and \mbox{$\alpha \simeq -1$} in Fig.~\ref{fig:NNN-magnetisierung-scanlog-um-1}. Clearly for $\alpha<-1$ and low temperature the magnetisation has a finite asymptotic limit for small magnetic fields, whereas this is not the case for $\alpha\!=\!-1$. So, a first order phase transition exists for $\alpha<-1$.
This statement also holds for $h=0$ and $\alpha>1$ and for the model with NN interactions with zero magnetic field and $|\alpha|>1$.

Finally, we determine the asymptotic behaviour of the specific heat for $T\!\rightarrow\! 0$ at the phase transitions. We find the specific heat for $h\!=\!0$ and $|\alpha|\!=\!1$ to very low temperatures and with very high accuracy.

\begin{figure}[t]
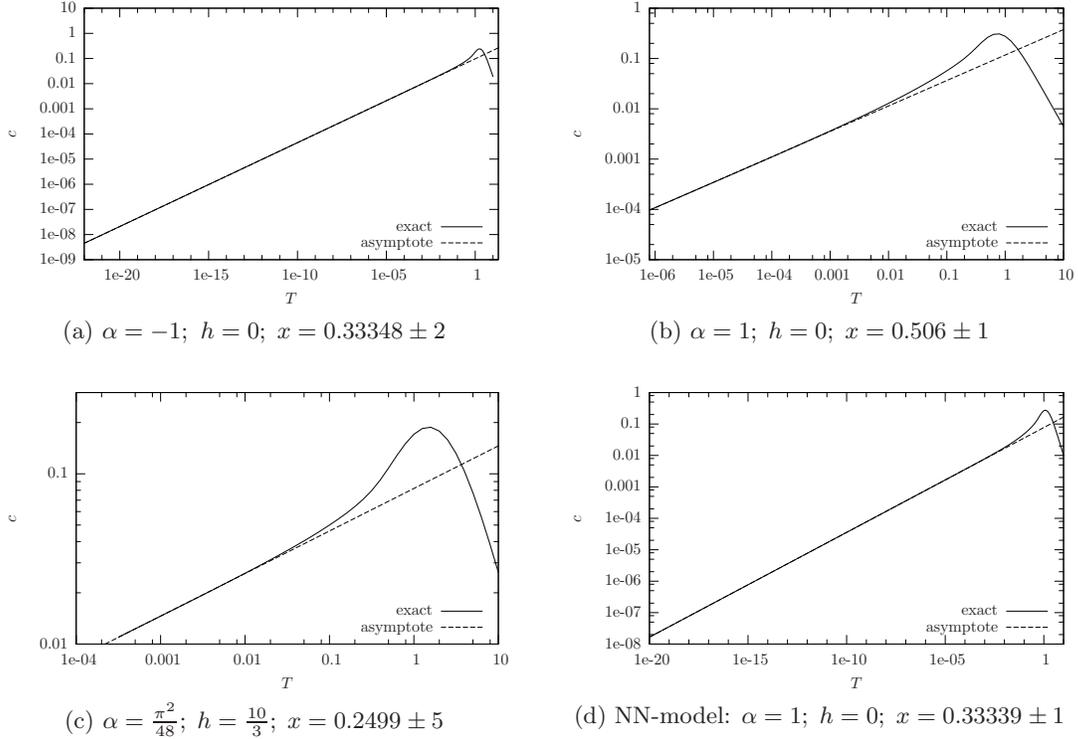

\centering
\subfloat[][$\alpha=-1;\ h=0;\; x=0.33348\pm2$]{\label{fig:NNN-spez-alpha-10} \resizebox{7cm}{!}{\input spez_NNN_a=-1,0_english.tex}}\quad
\subfloat[][$\alpha=1;\; h=0;\; x=0.506\pm1$]{\label{fig:NNN-spez-alpha10} \resizebox{7cm}{!}{\input spez_NNN_a=1,0_english.tex}}
\subfloat[][$\alpha=\frac{\pi^2}{48};\; h=\frac{10}{3};\; x=0.2499\pm5$]{\label{fig:NNN-tricritical} \resizebox{7cm}{!}{\input spez_NNN_trikritisch_english.tex}}\quad
\subfloat[][NN-model: $\alpha=1;\; h=0;\; x=0.33339\pm1$]{\label{fig:NN-spez-alpha10} \resizebox{7cm}{!}{\input spez_NN_a=1,0_english.tex}}
\caption{The specific heat for the model with NNN interactions (Fig.~\protect\subref{fig:NNN-spez-alpha-10}-\protect\subref{fig:NNN-tricritical}) and the model with NN interactions (Fig.~\protect\subref{fig:NN-spez-alpha10}). The low temperature exponent $x$ of $T$ ($c\sim T^x$) is determined by a fit on the numerical data. The error refers to the last digit and only includes the statistical error of the fit.}
\label{fig:spez-alpha1}
\end{figure}

The specific heat vanishes as $T^{1/3}$ for the model with NNN interactions at \mbox{$\alpha=-1$} (Fig.~\ref{fig:NNN-spez-alpha-10}) and as $T^{1/2}$ (Fig.~\ref{fig:NNN-spez-alpha10}) for $\alpha=1$.
For the phase transition into the ferromagnetic phase and with slightly lower numerical accuracy for the phase transition between the commensurate and incommensurate phase we find the asymptotic behaviour $T^{1/2}$.
At the tricritical point the specific heat vanishes like $T^{1/4}$ (see Fig.~\ref{fig:NNN-tricritical}) as predicted in \cite{ZvyKl:qpt}.
For the first order phase transitions we were not able to compute the specific heat at sufficiently low temperatures with sufficiently high accuracy to find consistent results for the low temperature asymptotics for $|\alpha|>1$. Here further investigations are necessary.

Again the situation in the model with NN interactions corresponds to the one in the model with NNN interactions and negative coupling $\alpha$. In particular this means that the specific heat at $\alpha=1$ and $h=0$ vanishes like $T^{1/3}$ which is shown in Fig.~\ref{fig:NN-spez-alpha10}.

\section{Conclusion}

We studied the thermodynamics and ground state phase diagrams of two integrable models containing  the standard spin-$\frac{1}{2}$ Heisenberg Hamiltonian and additional competing interactions.

The ground state phase diagrams depending on the external magnetic field and the longer range coupling $\alpha$ are constructed. They contain ferromagnetic and antiferromagnetic phases. In both models there exist second order phase transitions between these phases and first order phase transition lines with non-vanishing spontaneous magnetisation. Only the model with NNN interactions with positive $\alpha$ contains a phase transition between a commensurate and an incommensurate phase.

The NLIE describing the models at finite temperature are solved numerically for typical values of the coupling $\alpha$ and the magnetic field $h$.

The vicinity of the phase coexistence in both models  is difficult to investigate. For sufficiently low temperatures the NLIE are numerically ill-posed if straight integration contours are used. For reaching low temperatures, either the contours have to be deformed - or as chosen in our approach - the truncation level has to be increased.

\section*{Acknowlegment}
The authors like to acknowledge support by the research program of the Graduiertenkolleg
1052 funded by the Deutsche Forschungsgemeinschaft.

%

\begin{thebibliography}{10}

\bibitem{GK:hubbard-model}
F.H.L. Essler, H.~Frahm, F.~G\"ohmann, A.~Kl\"umper, and V.E. Korepin,
  \emph{The one-dimensional {H}ubbard model}, Cambridge, 2005.

\bibitem{Frahm:xxz}
H.~Frahm, \emph{Integrable spin-1/2 {XXZ} {H}eisenberg chain with competing
  interactions}, J.~Phys.~A: Math.~Gen. \textbf{25} (1992), 1417--1427.

\bibitem{FrahmRoed:spinladder99}
H.~Frahm and C.~R\"odenbeck, \emph{A generalized spin ladder in a magnetic
  field}, Eur.~Phys.~J.~B \textbf{10} (1999), 409--414.

\bibitem{Kluemper92}
A.~Kl\"umper, \emph{Free energy and correlation lengths of quantum chains
  related to restricted solid-on-solid lattice models}, Ann.~Phys. \textbf{1}
  (1992), 540--553.

\bibitem{Kluemper93}
\bysame, \emph{Thermodynamics of the anisotropic spin-1/2 {H}eisenberg chain
  and related quantum chains}, Z.~Phys.~B \textbf{91} (1993), 507--519.

\bibitem{KlSa:thermal}
A.~Kl\"umper and K.~Sakai, \emph{The thermal conductivity of the spin-1/2 {XXZ}
  chain at arbitrary temperature}, J.~Phys.~A: Math.~Gen. \textbf{35} (2002),
  2173--2182.

\bibitem{KBI:qism}
V.E. Korepin, N.M. Bogoliubov, and A.G. Izergin, \emph{Quantum {I}nverse
  {S}cattering {M}ethod and {C}orrelation {F}unctions}, Cambridge, 1997.

\bibitem{MuellerVekuaMikeska2002}
M.~M\"uller, T.~Vekua, and H.J. Mikeska, \emph{Perturbation theories for the
  $s=\frac{1}{2}$ spin ladder with a four-spin ring exchange}, Phys.~Rev.~B
  \textbf{66} (2002), no.~13, 134423.

\bibitem{MuramotoTakahashi:four-body}
N.~Muramoto and M.~Takahashi, \emph{Integrable {M}agnetic {M}odel of {T}wo
  {C}hains {C}oupled by {F}our-{B}ody {I}nteractions}, J.~Phys.~Soc.~Jpn.
  \textbf{68} (1999), no.~6, 2098--2104.

\bibitem{Su:fh}
J.~Suzuki, \emph{Spinons in magnetic chains of arbitrary spins at finite
  temperatures}, J.~Phys.~A: Math.~Gen. \textbf{32} (1999), 2341--2359.

\bibitem{Takahashi:TBA}
M.~Takahashi, \emph{One-{D}imensional {H}eisenberg {M}odel at {F}inite
  {T}emperature}, Prog.~Theor.~Phys. \textbf{46} (1971), no.~2, 401--415.

\bibitem{Tsvelik:mxxz}
A.M. Tsvelik, \emph{Incommensurate phases of quantum one-dimensional
  magnetics}, Phys.~Rev.~B \textbf{42} (1990), no.~1, 779--785.

\bibitem{zvyagin:BA-multichain}
A.A. Zvyagin, \emph{Bethe ansatz solvable multi-chain quantum systems},
  J.~Phys.~A: Math.~Gen. \textbf{34} (2001), R21--R53.

\bibitem{ZvyKl:qpt}
A.A. Zvyagin and A.~Kl\"umper, \emph{Quantum phase transitions and
  thermodynamics of quantum antiferromagnets with next-nearest-neighbor
  couplings}, Phys.~Rev.~B \textbf{68} (2003), 144426.

\end{thebibliography}

\providecommand{\bysame}{\leavevmode\hbox to3em{\hrulefill}\thinspace}
\providecommand{\MR}{\relax\ifhmode\unskip\space\fi MR }
\providecommand{\MRhref}[2]{%
  \href{http://www.ams.org/mathscinet-getitem?mr=#1}{#2}
}
\providecommand{\href}[2]{#2}

\end{document}